\documentclass[11pt,twoside]{article}

\usepackage{asp2006}
\usepackage{epsf}
\usepackage{psfig}
\usepackage{lscape}
\usepackage{graphicx}
\usepackage{subfigure}

\begin{document}
\title{Massive Young Stars in the Galactic Center}   

\author{ 
H.~Bartko\altaffilmark{1},
F.~Martins\altaffilmark{1,2},  
S.~Trippe\altaffilmark{10} 
T.~K.~Fritz\altaffilmark{1},  
R.~Genzel\altaffilmark{1,3},  
T.~Ott\altaffilmark{1},  
F.~Eisenhauer\altaffilmark{1},  
S.~Gillessen\altaffilmark{1},  
T.~Paumard\altaffilmark{5},  
T.~Alexander\altaffilmark{6}, %
K.~Dodds-Eden\altaffilmark{1},  
O.~Gerhard\altaffilmark{1}, %
Y.~Levin\altaffilmark{4}, %
L.~Mascetti\altaffilmark{1},  
S.~Nayakshin\altaffilmark{7}, %
H.~B.~Perets\altaffilmark{8}, %
G.~Perrin\altaffilmark{5}, %
O.~Pfuhl\altaffilmark{1},  
M.~J.~Reid\altaffilmark{8},  
D.~Rouan\altaffilmark{5}, %
M.~Zilka\altaffilmark{9}, 
A.~Sternberg\altaffilmark{9} 
}
 

\begin{abstract} 
We summarize our latest observations of the nuclear star cluster in the central parsec of the Galaxy with the adaptive optics assisted, integral field spectrograph SINFONI on the ESO/VLT, which result in a total sample of 177 bona fide early-type stars.
We find that most of these Wolf Rayet (WR), O- and B- stars reside in two strongly warped eccentric ($<e> = 0.36\pm0.06$) disks between 0.8'' and 12'' from SgrA*, as well as a central compact concentration (the S-star cluster) centered on SgrA*. The later type B stars ($m_K>15$) in the radial interval between 0.8'' and 12'' seem to be in a more isotropic distribution outside the disks. We observe a dearth of late-type stars in the central few arcseconds, which is puzzling.
The stellar mass function of the disk stars is extremely top-heavy with a best fit power law of $\mathrm{d}N/\mathrm{d}m \propto m^{-0.45\pm0.3}$.
Since at least the WR/O-stars were formed in situ in a single star formation event $\sim$6 Myrs ago, this mass function probably reflects the initial mass function (IMF). The mass functions of the S-stars inside 0.8'' and of the early-type stars at distances beyond 12'' differ significantly from the disk IMF; they are compatible with a standard Salpeter/Kroupa IMF (best fit power law of $\mathrm{d}N/\mathrm{d}m \propto m^{-2.15\pm0.3}$).
\footnotetext[1]{Max-Planck-Institute for Extraterrestrial Physics, Garching, Germany\\
$^{}$ \hspace{-0.4cm} $^2$GRAAL-CNRS, Universit Montpelier II - UMR5024, Montpellier, France\\
$^{}$ \hspace{-0.4cm} $^3$Department of Physics, University of California, Berkeley, USA\\
$^{}$ \hspace{-0.4cm} $^4$Leiden University, Leiden Observatory and Lorentz Institute, , Leiden, the Netherlands\\
$^{}$ \hspace{-0.4cm} $^5$LESIA, Observatoire de Paris, CNRS, UPMC, Université Paris Direrot, Meudon, France\\
$^{}$ \hspace{-0.4cm} $^6$Faculty of Physics, Weizmann Institute of Science, Rehovot 76100, Israel \\
$^{}$ \hspace{-0.4cm} $^7$Department of Physics \& Astronomy, University of Leicester, Leicester, UK\\
$^{}$ \hspace{-0.4cm} $^8$Harvard-Smithsonian Center for Astrophysics, 60 Garden Street, Cambridge, USA\\
$^{}$ \hspace{-0.4cm} $^9$School of Physics and Astronomy, Tel Aviv University, Tel Aviv 69978, Israel\\
$^{}$ \hspace{-0.4cm} $^{10}$IRAM Grenoble, F-38406 Saint Martin d'Heres, France.
}
\end{abstract}

\vspace{-0.3cm}
\section{Introduction}   


The central parsec of the Galaxy harbors more than one hundred young massive stars \citep[see][and references therein]{Paumard2006,Bartko2009,Bartko2010}. 
This is highly surprising, since the tidal forces from the central four million solar mass black hole should make formation of stars by gravitational collapse from a cold interstellar cloud very difficult, if not impossible \citep{Morris1993}. 
Most of the WR-stars and O-stars (dwarfs, giants, and supergiants) dominating the luminosity of the early-type population were formed in a well defined single event $\sim$6 Myrs ago, perhaps as the result of the infall of a gas cloud followed by an in-situ star formation event. About half of these WR/O-stars between 0.8'' and 12'' from SgrA* reside in a well defined but highly warped and eccentric disk that rotates clockwise on the sky. 
There exists a significant ($7.1\sigma$) counter-clockwise structure of WR/O stars \citep{Paumard2006,Bartko2008,Bartko2009,Bartko2010}, perhaps a second disk in a dissolving state. 
Within about 1'' of SgrA* there is a sharp cutoff in the density of WR/O-stars. Instead there is a concentration of fainter stars (B dwarfs) with randomly oriented and eccentric orbits: the so-called 'S-star cluster' \citep{Ghez2003,Gillessen2009}. 
 

\vspace{-0.3cm}
\section{Results}   

\begin{figure}[t!] 
\begin{center} 
\includegraphics[totalheight=5.6cm]{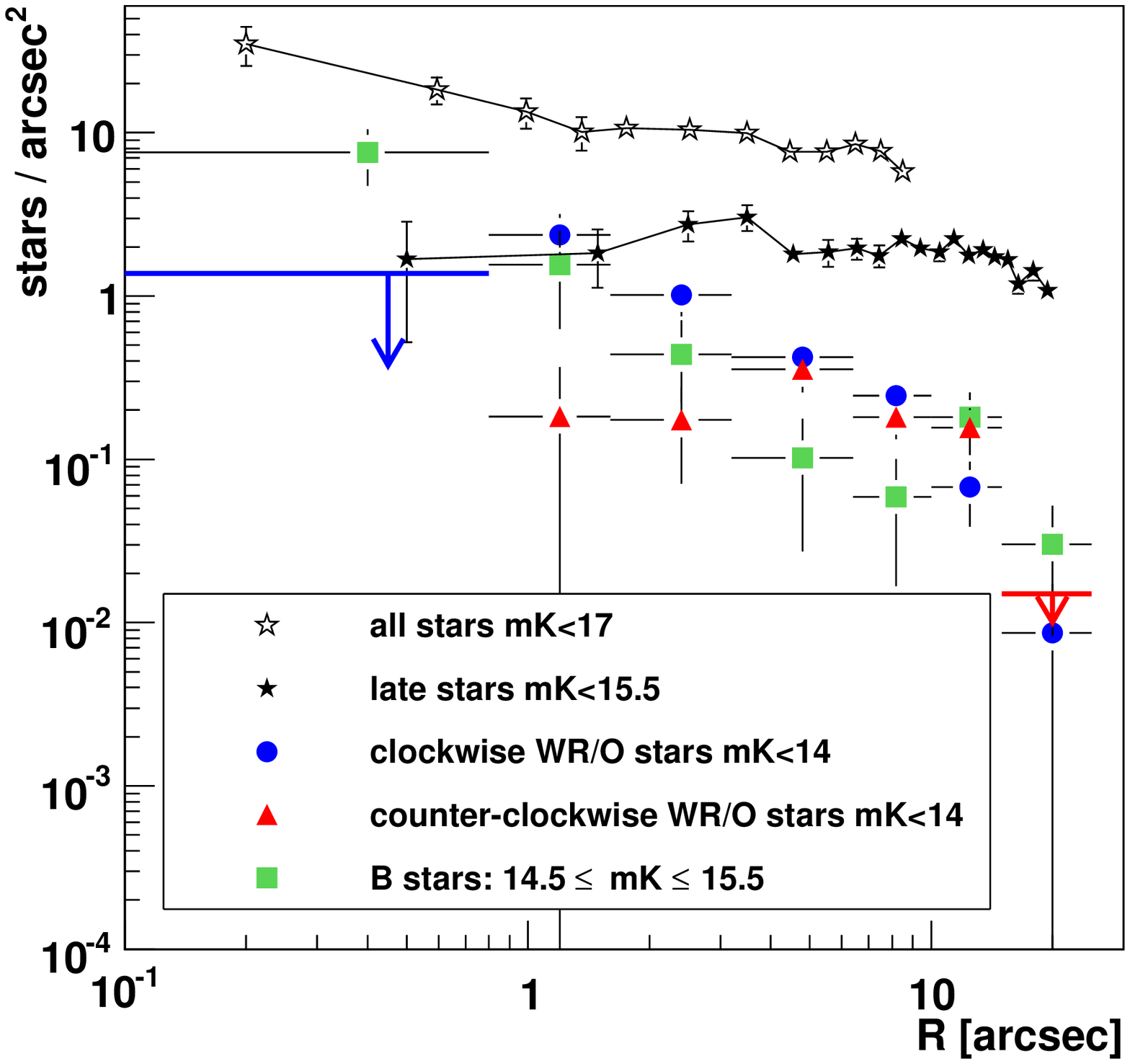}
\includegraphics[totalheight=6cm]{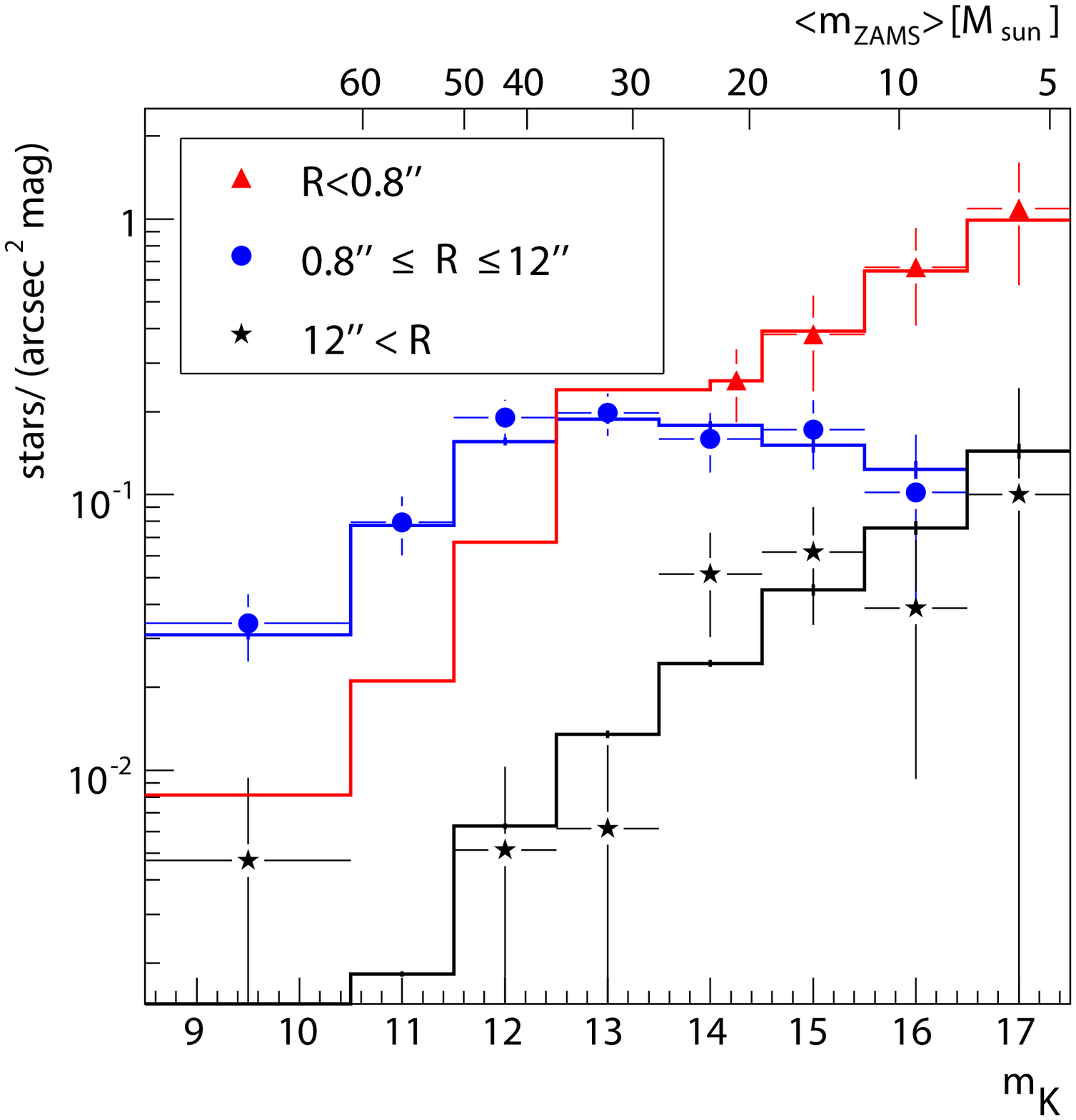}
\vspace{-0.3cm}
\caption{\it \small (Left) Projected, completeness and coverage corrected surface density of different stellar populations as a function of distance to SgrA*. 
(Right) Completeness corrected K-band luminosity functions of early-type stars in three radial intervals $R<0.8''$ (red triangles, scaled by a factor 0.05), $0.8'' \leq R \leq 12''$ (blue points) and $12''<R<25''$ (black asterisks).
} 
\vspace{-0.5cm}
\label{fig1} 
\end{center} 
\end{figure}

Figure \ref{fig1} (left) shows the projected, completeness and coverage corrected surface density profiles for various stellar populations.  
The radial surface density of the clockwise WR/O-stars follows a power law $\Sigma(R) \propto R^{-1.4\pm0.2}$ between 0.8'' and 15''. 
The surface density distribution of the counter-clockwise WR/O-stars is flat between 0.8'' and 15'' with a larger central 'hole' than the one of the clockwise stars and a shallow maximum at around 4.5''.  
The radial surface density profile of B-dwarfs in the magnitude interval $14.5 \leq m_K \leq 15.5$ drops smoothly 
from the central 0.8'', where the S-stars reside, out to about 25''.  
The distribution is similar to the clockwise WR/O-stars with a best-fitting power law of $\Sigma(R) \propto R^{-1.5\pm0.2}$. In strong contrast to the early-type stars, late-type stars with $m_K \leq 15.5$ exhibit a flat surface density distribution inside of 10'' \citep[see also][]{Buchholz2009,Do2009}.

Figure \ref{fig1} (right) shows the completeness corrected K-band luminosity functions of early-type stars in three radial intervals 
together with three theoretical model luminosity functions based on different IMFs, computed with the population synthesis code STARS \citep{Sternberg2003}. We assumed a cluster age of 6 Myrs, and an exponentially decaying star burst with an $1/e$ time-scale of 1 Myrs. This is the best-fitting age and duration of a single star formation event derived from the Hertzsprung-Russell diagram distribution of the O/B-stars and the ratios of various sub-types of WR-stars \citep{Paumard2006}. 
The KLF in the radial interval $0.8'' \leq R \leq 12''$ can be fitted by a power-law IMF of $\mathrm{d}N/\mathrm{d}m \propto m^{-0.45\pm0.3}$ (the Salpeter IMF is  $\mathrm{d}N/\mathrm{d}m \propto m^{-2.3}$). 
The IMF found by \citet{Bonnell2008} for their $10^5 M_{\odot}$ cloud simulation fits our data reasonably well. 
The IMF of the field early-type stars beyond 12'' and the S-stars within 0.8'' can be fitted by $\mathrm{d}N/\mathrm{d}m \propto m^{-2.15\pm0.3}$. 
These latter KLFs can also be fitted by Salpeter/Kroupa IMFs and continuous star formation histories with moderate ages ($\leq 60$ Myrs). 
 
Most of the stars with $m_K>14$ are B dwarfs, the brighter magnitude bins contain O dwarfs, giants and WR stars. The main sequence ends at about $m_K=12$, corresponding  to stars with initial masses of about $25 M_{\odot}$ with main-sequence lifetimes of 6 Myr.
An $m_K=16.5$ early-type star for a 6 Myr old population corresponds approximately to a B5V main sequence star with a ZAMS mass of about $7 M_{\odot}$ (see Figure \ref{fig1} right). The most massive stars with individual mass estimates have ZAMS masses of at least $60 M_{\odot}$ \citep{Ott1999,Martins2006,Martins2007}. 
Our estimated IMF slope of the stellar disks ($\mathrm{d}N/\mathrm{d}m \propto m^{-0.45\pm0.3}$) is therefore at least valid over the mass interval $7-60 M_{\odot}$.


50 out of 110 stars with $m_K<14$ have an angular offset to the local average angular momentum direction of the clockwise system \citep{Bartko2009}, below $20^{\circ}$ and 8 out of 15 stars with  $14 \leq m_K \leq 15$ have offsets below $20^{\circ}$. In contrast, only one out of 11 stars fainter than $m_K=15$ is compatible with the clockwise disk. This may be an indication that most of the few later B-dwarfs observed in the region of the disks do not belong to the clockwise system, but rather to the background population seen at projected distances beyond 12'', which has a more standard IMF.

\vspace{-0.2cm}
\section{Discussion and Conclusions} 
\label{disc} 

The observations suggest that the two WR/O/B-star disks (or planar sets of streamers) between 0.8'' and 12'' are structurally and dynamically distinct from both the S-star cluster and the outer region. The properties of these warped disks, including in particular their steep surface density distribution, their top-heavy IMF \citep[see also][]{NayakshinSunyaev2005}, and their relatively low total stellar masses all strongly favor an in situ star formation model \citep{Paumard2006, Lu2009, Bartko2009}
and broadly agree with the findings of recent hydrodynamical simulations \citep{Bonnell2008,Hobbs2009} of star formation triggered by infalling gas clouds. The sharp transition at 0.8'' between disk zone and S-star cluster, both in terms of dynamics and mass function, in our opinion also strongly disfavors migration scenarios from the disks for the origin of the S-star cluster.
In turn, this sharp transition supports 'injection and capture' scenarios \citep[e.g.][]{Hills1988,Perets2007}. 

The large central core \citep[][]{Buchholz2009,Do2009,Bartko2010} 
of late-type stars is puzzling and currently not understood. There are a number of possible interpretations, none of which at present offers a compelling explanation: Equilibrium mass segregation by itself cannot account for this distribution \citep{Freitag2006,HopmanAlexander2006}. 
Physical collisions with main-sequence stars and stellar black holes can remove moderately bright giants in 
the central arcsecond, but not over a much larger region, nor to $m_K \sim 15$ \citep{Dale2009}. 
Tidal disruption of stars may play a role in the immediate vicinity of the massive black hole \citep{Perets2009}, but most likely cannot be an explanation
for the lack of an old cusp on the ~0.5 pc scale. 
The in-spiral of an intermediate mass
black hole can plausibly gouge out a large enough core in the stellar distribution \citep{MilosavljevicMerritt2001,Baumgardt2006} but no direct evidence for such a second black hole was found \citep{Reid2004,Trippe2008,Gillessen2009,GualandrisMerritt2009}. 
If the two-body relaxation time scale is significantly longer than the age of the Galactic Center star cluster (and the Hubble time) throughout the central parsec \citep{Merritt2009} 
the observed large core may reflect the initial conditions of the Galactic Center nuclear cluster, but the dynamics of the old star cluster is consistent with a relaxed, fully phase-mixed system \citep{Trippe2008}. In addition, there may be relaxation processes much faster than the standard two-body rate \citep{Alexander2007,Perets2007}. The top-heavy IMF discussed here (if applicable to earlier star formation episodes) would obviously also lead to a lack of old low mass giants in the core, but the IMF probably would have to depend strongly on radius.
The remarkable properties of the Galactic Center nuclear
star cluster remain puzzling and continue to give us food for thought and further work.

 
 



\end{document}